# Multiplexed entangled photon sources for all fiber quantum networks

Yin-Hai Li[1, #], Zhi-Yuan Zhou[2,3, #], Zhao-Huai Xu[1], Li-Xin Xu[1, *], and Bao-Sen Shi[2,3, +], Guang-Can Guo[2,3]

[1]*Department of Optics and Optical Engineering, University of Science and Technology of China, Hefei, Anhui 230026, China*
[2]*Key Laboratory of Quantum Information, University of Science and Technology of China, Hefei, Anhui 230026, China*
[3]*Synergetic Innovation Center of Quantum Information & Quantum Physics, University of Science and Technology of China, Hefei, Anhui 230026, China*

*[*xulixin@ustc.edu.cn](mailto:xulixin@ustc.edu.cn)*
*[+drshi@ustc.edu.cn](mailto:drshi@ustc.edu.cn)*

The ultimate goal of quantum information science is to build a global quantum network, which enables quantum resources to be distributed and shared between remote parties. Such quantum network can be realized by all fiber elements, which takes advantage of low transmission loss，low cost, scalable and mutual fiber communication techniques such as dense wavelength division multiplexing (DWDM). Therefore high quality entangled photon sources based on fibers are on demanding for building up such kind of quantum network. Here we report multiplexed polarization and time-bin entanglement photon sources based on dispersion shifted fiber (DSF) operating at room temperature. High qualities of entanglement are characterized by using interference, Bell inequality and quantum state tomography. Simultaneous presence of entanglements in multi-channel pairs of a 100GHz DWDM shows the great capacity for entanglements distribution over multi-users. Our research provides a versatile platform and moves a first step toward constructing an all fiber quantum network.

Entanglement is one of the key resources in quantum information processing, which enables some incredible applications such as quantum teleportation [1-3], dense coding [4], enhanced sensitivity in image and metrology [5-7]. Usually, entanglement can be created in various physical systems based on second or third order nonlinear processes, some instances of entanglement generation include: spontaneous parametric down conversion (SPDC) based on bulk or waveguide nonlinear crystals [8, 9]; spontaneous Raman scattering(SRS) or spontaneous four wave mixing (SFWM) in atomic ensembles [11, 12], dispersion shifted fibers (DSF)[13-15], silicon waveguide [16-18] and photonic crystal fibers(PCF)[19]. For conserving of energy, linear momentum and angular momentum in these nonlinear interaction processes, entanglement can be created in different degree of freedoms of photon, the common generated photonic entangled source are polarization, time-bin, path, and orbital angular momentum entanglement[20].

---

[#]*These two authors contributed equally to this work.*

Though various system can be used for preparing different kinds of entanglements, for the purpose of constructing a global quantum network and distributing entanglement over remote parties, to generate photonic entanglements lie in the low loss transmission windows of fiber is indispensable [21]. To generate entanglement photon source using commercial DSF shows evident advantages over other methods for its directly connectable to fiber, free of free space alignment, a large amount of cheap fiber elements can be used at your hand, easily to be integrated for scaling and some mutual fiber communication techniques such as dense wavelength division multiplexing (DWM) and time division multiplexing can be used to enhance the information transmission capacity. In the past decades, big advances in preparing polarization and time-bin entangled photon sources are made by using DSF [13-15] or photonic crystal fiber [19]. Also significant progresses in suppressing the intense Raman scattering noise of photon pair generating in DSF have been achieved by using pulse pumping, polarization filtering, high performance superconducting nanowire single photon detectors for time filtering and cooling with liquid gasses [22, 23], the coincidence to accidental coincidence ratio is increased to very high level, which is feasible to be used in quantum information processing task. Very recently, ground break works on quantum storage of telecom band photonic polarization qubit, time-bin entangled state and frequency multiplexed modes in erbium doped optical fiber are demonstrated [24-26], which show great potential for implementing a long distance all fiber quantum networks. All fiber entangled photon sources are good candidate for achieving this goal. Though there are many works reported on preparing entangled photon source based on DSF, no one reported wavelength multiplexed entangled source based on DSF, which has the same figure of merits as reported in ref. [18], these merits include: realizing entanglement over multi-frequency modes, which enable engineering a complex quantum state; compatible with contemporary telecom fiber and quantum memories, and with chip-scale semi-conductor technology, which is capable for compact, low cost and scalable applications.

In this work, we show multiplexed entanglement photon sources over three pairs of 100GHz DWDM channels for both polarization and time-bin based on DSF for the first time. Our source has good entanglement quality operating at room temperature, the coincidence to accidental coincidence ratio (CAR) reaches 30 without cooling the DSF with liquid gases. The visibilities are nearly 90% without subtraction of background coincidences, and reaches near unity after the background is subtracted. The CHSH equality measured for the polarized entangled source has S value of 2.38 ±0.12 with violation of more than 3 standard deviations. By quantum state tomography for the time-bin entangled source, we obtain a fidelity of 0.9120±0.0120 (0.9723±0.0070)without (with) background subtraction. All three channel pairs have high two-photon interference visibilities, which show the capability to distribute entanglement over multi-users. The present multiplexed entangled photon sources are of great importance for realizing an all fiber quantum networks and enhancing the transmission capacity by using the state-of-art DWDM technique.

**Multiplexed polarization entangled source**

The experimental setup for generating polarized entangled photon source is showed in figure 1. A homemade mode locked fiber laser centered at 1550.1nm, which has repetition rate of 27.9MHz and pulse width of 25 ps, is used as the pump beam []. The pump beam is firstly passed through a variable attenuator for adjusting the pump power, then the broad band background fluorescence photon is filtered out with cascading of 100GHz DWDM filters. The clean pump beam is used to pump a fiber loop with 300 m DSF for generating polarization entangled photon pairs through SFWM. The strong pump beam is removed by cascading 200GHz DWDM filters, then a 32 Channel 100GHz DWDM is used for distributing polarization entanglement at correlated channel pairs. Output ports of each channel pairs are connected to avalanched photon detector (APD1, APD2, IDQ, ID220), output signals from ADP1, 2 are sent to our coincidence count device (Pico quanta, timeharp 260, 0.4 ns coincidence window).

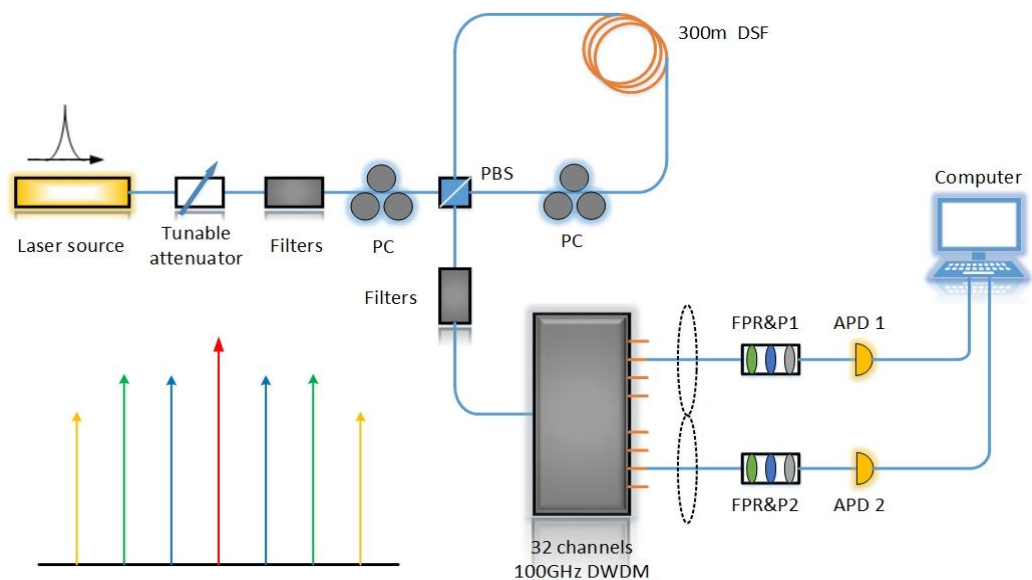

Figure 1. Experimental setup for multiplexed polarization entangled photon pairs.

The principle for this fiber loop is similar to a Sagnac interferometer loop for generating polarization entangled photon source based on second order nonlinear crystals in the SPDC processes [9]. In the present case, the pump beam is split into clockwise and counterclockwise directions by a polarized beam splitter (PBS) for generation photon pairs with polarization states of $|H\rangle_s|H\rangle_i$ and $|V\rangle_s|V\rangle_i$. After photon pairs from the two counter propagation directions recombined at the PBS, the photonic state at the output port of the fiber loop can be expressed as [14]

$$|\Phi\rangle=|HH\rangle+\eta e^{i\delta}|VV\rangle \quad (1)$$

Where $\eta$ is determined by the ratio of pump power; $\delta=2(\varphi_p+\Delta k_p L)$ depends on the initial phase $\varphi_p$ of the pump beam at the input port and the birefringence

$\Delta k_p = k_{pH} - k_{pV}$ experienced by the pump beam in the H and V polarizations. By changed the pump power and initial phase, we can generate the maximally entangled states $|\Phi\rangle^{\pm} = 1/\sqrt{2}(|HH\rangle \pm |VV\rangle)$.

In order to characterize the quality of our entangled source, different methods are used in our experiment. We firstly measure the two-photon polarization interference fringes at different settings. The experimental result is showed in figure 2(a), the raw (net) visibilities in the 0 and 45 degree basis for channel C31-C37 are 89.33±1.91% (96.61±1.06%) and 87.92±1.95% (94.39±1.33%) respectively. The greater than 71% visibilities imply the presence of quantum entanglement.

To further characterize the performance of the entangled source, we measure the CHSH inequality S parameter [27]. The S parameter measured for two sets of polarization direction settings ($\theta_s = -22.5^0, 67.5^0, 22.5^0, 112.5^0$; $\theta_i = -45^0, 45^0, 0^0, 90^0$) is 2.38±0.12 without background subtracted, which violates the inequality with more than 3 standard deviations. After subtracting the background, the S parameter is 2.50±0.12.

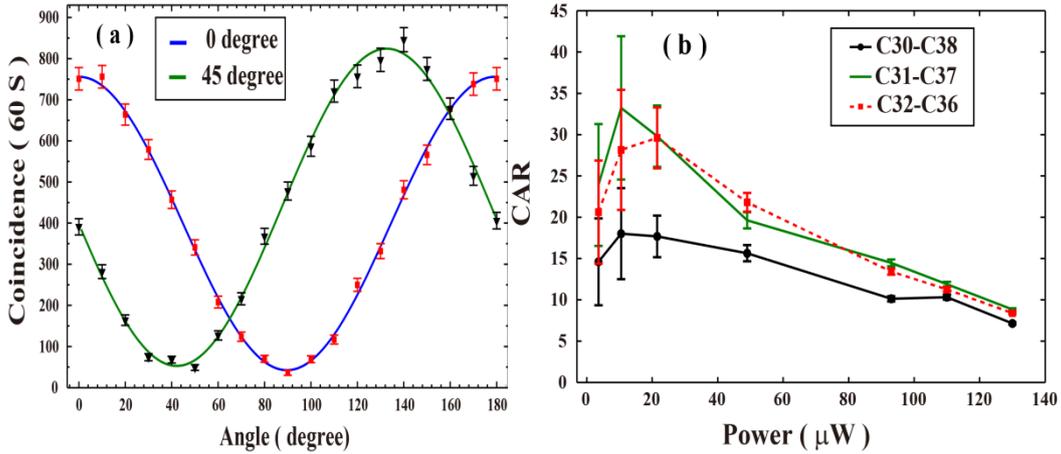

Figure 2. Two photon interference fringes and CARs for multiplexed polarization entangled source. (a)Coincidences in 60 s for channel pair C31-C37 as function of signal polarizer angle, when the idler angle is fixed at 0 and 45 degree. (b)CARs for different correlated channel pairs as function of average pump power.

To show the capability of entanglement distribution over multi-users by using wavelength multiplexing, we measurement the CARs over three correlated channel pairs by varying the pump power. The result is depicted in figure 2(b), the maximum CAR reaches 30 at the average pump power of 20 μW. Channel pairs C31-C37 and C32-C36 have similar behaviors, but channel pair C30-C38 has relative lower CAR because of the spectral of photons in channel C30 and C38 are near the boundary of the emitted photon pairs. The high CAR at room temperature demonstrates the abilities for polarization entanglement distribution over multi-users by using DWDM technique.

To know the actual performance of our multi-channel polarization entangled photon pairs, we also measured the visibilities in 0 and 45 basis for the three channel pairs at average pump power of 100 μW. The results are listed in table 1. All the raw visibilities are greater than 80%, and the net visibilities are greater than 96%. The simultaneous presence of high visibilities implies the high quality of our multi-channel polarized entangled photon source. The average losses excluding the detectors efficiency for signal and idler photon is around 8 dB, therefore the photon pairs generation rates in 100GHz bandwidth of the channel pair C31-C37 is 1MHz, which is equal to 0.11 Hz per pulse.

Table. 1. Visibilities for different channel pairs for 100 μW pump and 30 S coincidences

| Channel | Raw V0 ( % ) | Net V0 ( % ) | Raw V45 (%) | Net V45 (%) |
| --- | --- | --- | --- | --- |
| C31-C37 | 90.86±1.76 | 98.34±0.72 | 90.50±1.70 | 97.07±0.93 |
| C32-C36 | 89.85±1.50 | 98.91±0.47 | 87.05±1.73 | 95.96±0.96 |
| C30-C38 | 80.87±3.11 | 98.25±0.93 | 82.79±2.71 | 97.48±1.03 |

**Multiplexed time-bin entangled source**

Time-bin entangled photon pair source is another vital important quantum resource, which is suitable for quantum key distribution over telecom fibers. The scheme for time-bin entanglement generation is shown in figure 3, the pump laser, forward and backward filters are the same as previous experiment. The pump beam is split into two time bins by using a 1.6 ns unbalanced Michelson fiber interferometer (UMI) [28,29], which is consist of one fiber coupler and two Faraday rotator mirrors. Each pump time-bin generated a pair of photon pair, after removing the pump beam, the generated photon pairs are separated by a 32 Channel 100GHz DWDM. The correlated channel pairs are connected to two UMIs with the same parameters as the first UMI. Each UMI is separately sealed in an copper box and thermal insulation from the air, the temperature of the copper boxes are controlled with homemade semiconductor Peilter temperature controller with temperature fluctuation of ±2mK. Output of one of the interferometer is connected to an APD1(lightwave Princeton, 100MHz trigger rate, 1 ns detection window, 15% detection efficiency), which is gated by the synchronize signal from the mode-locked laser. The output from another interferometer is connected to APD2 (IDQ, ID220, free running detection, 20% detection efficiency, 10 us dead time). The electrical detection output signal from APD1 and APD2 are sent for coincidence measurement.

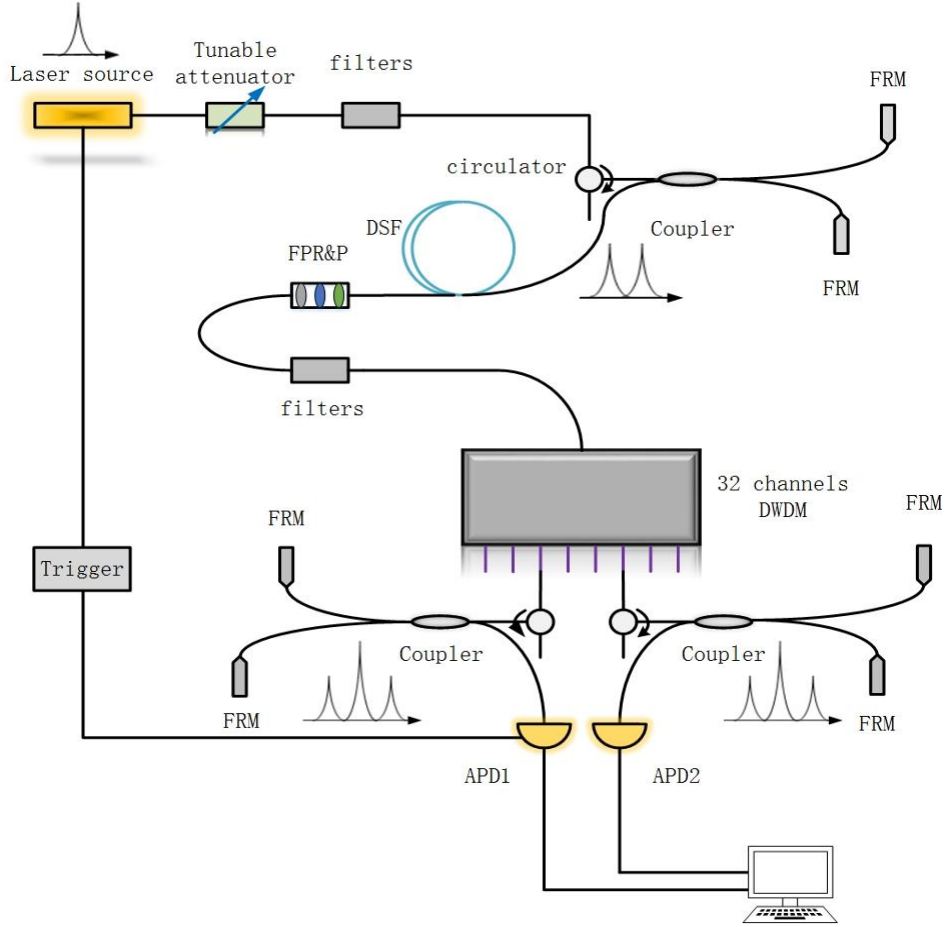

Figure 3. experimental setups for multiplexed time-bin entangled photon pairs.

First of all, we'll give a brief theoretical description of our time-bin entanglement source. After the pump beam is divided into two time slots by forward UMI, the pump photon is converted to state $|\Psi\rangle_p = 1/\sqrt{2}(|S\rangle - e^{i\phi_p}|L\rangle)$, where $|S\rangle$ and $|L\rangle$ denote the photons pass through the short and long arms of the UMI respectively, $\phi_p$ is the relative phase between the two arms. After transmitting of the pump beam from the DSF, time-bin entanglement is created from the two time slots, which can be expressed as [15]

$$|\Phi\rangle = \frac{1}{\sqrt{2}}(|SS\rangle - e^{i2\phi_p}|LL\rangle) \quad (2)$$

When the entanglement state in equation (2) is further transformed by two UMIs in the signal and idler ports, and we post-select the central slot, we will obtain two-photon interference fringes. The coincidence of the two photon is proportional to $1 - V\cos(2\phi_p - \phi_s - \phi_i)$, here $\phi_s$ and $\phi_i$ are the relative phase of the UMIs in the signal and idler ports respectively.

To characterize the quality of our time-bin entangled source in detail. We first measure dependences of the two photon interference with respected to the pump, signal and idler UMIs phases ($\phi_p, \phi_s, \phi_i$) by tuning the temperatures of the three UMIs. The results are showed in figure 4(a) and 4(b). In figure 4(a), the pump phase is kept at 0, the idler phase is kept at 0 and π/4 for the red solid and green dashed curves respectively. The raw (net) visibilities are 90.45±2.30% (98.58±0.85%) and 87.65±2.59% (95.14±1.62%) respectively. Two curves in figure 4(b) show the results when $\phi_p = \phi_i = 0$ and $\phi_s = \phi_i = 0$, which verify that the two photon interference fringe has oscillation period of π for pump phase and 2π for signal (idler) phase.

To fully characterize a quantum state, quantum state tomography is required. We use the method introduced in ref. [30] for projected measurements on different basis. The real and imaginary part for our reconstructed density matrix are showed in figure 4(c) and 4(d), the raw (net) fidelity of our experimental density matrix from the ideal density matrix is $0.9120 \pm 0.0120$ ($0.9723 \pm 0.0070$). High fidelity of the reconstructed density matrix shows the good quality of our time-bin entangled source.

To show the feasible of distribution our time-bin entangled source over multi-channels, we measured the visibilities of the three correlated channel pairs, the results are listed in table 2, the raw visibilities are greater than 84% and the net visibility is greater than 95%, which implies that all the channel pairs have high entanglement quality, and suitable for all fiber distribution of time-bin entanglement to multi-users by using DWDM technique.

Table. Visibilities for different channel pairs for 100 uW pump and 30 S coincidences

| Channel | Count(APD1, k/s) | Count(APD2, k/s) | Raw visibility (%) | Net visibility (%) |
| --- | --- | --- | --- | --- |
| C31-C37 | 5.8 | 18.0 | 89.42±2.35 | 96.85±1.26 |
| C32-C36 | 6.4 | 20.0 | 87.35±2.41 | 97.74±0.99 |
| C30-C38 | 5.0 | 19.0 | 83.98±3.65 | 95.24±1.99 |

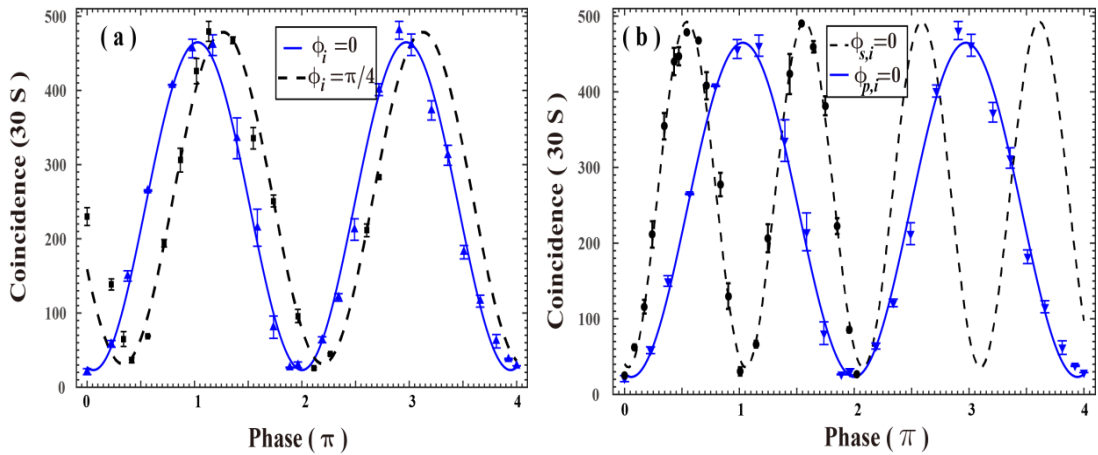

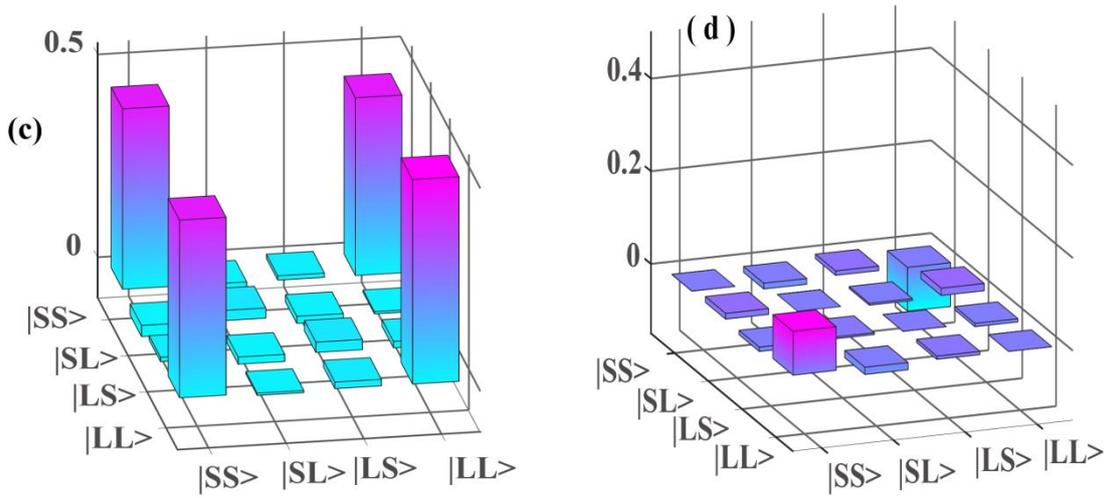

Figure 4. Two photon interference fringes and reconstructed density matrix for multiplexed time-bin entangled photon pairs. (a) Coincidences in 30 s as function of phase of signal UMI, when the pump UMI phase is fixed at 0, and the idler UMI phase is fixed at 0 and $\pi/4$, respectively. (b)Coincidences in 30 s when pump and idler UMIs phases are fixed at 0 and signal and idler UMIs phases are fixed at 0, respectively. (c), (d) Real and imaginary parts of our experimental reconstructed density matrix.

**Discussion and conclusions**

Polarization entangled and time-bin entangled photon source are two kinds of most important quantum resources for quantum computation and quantum communications. In this work, we prepare both kinds of entangled source based on SFWM in DSF at room temperature, the highest CAR is reached for room temperature operation of DSF based photon pair source. We also verify simultaneous presences of high entanglement over three channel pairs and the capability to use our source to distribute both kind of entanglement to multi-users by using DWDM technique, which is reported for the first time for DSF based photon pair source.

Though relative high performances are obtained for our source, further improvements of our DWDM entangled sources are feasible. In the present demonstration the number of channel pairs is limited to three, by engineering the dispersion properties of the DSF or using high nonlinear fiber with shorter fiber length, the number of channel pairs will be greatly extended. The intense Raman scattering noise for room temperature operation of DSF photon source limited the further increasing of CAR, by cooling the fiber to liquid nitrogen (77K) or helium (4K) temperature, the noise photon will be greatly reduced. The APDs used in our experiments have relative low efficiency, large time jitters and long dead time, by using high performance superconductor single photon detectors, the coincidence count rate will be greatly enhanced and the CAR will be further improved by apply a narrower coincidence window.

Quantum information processing based on integrated optical elements such as optical fibers and silicon on insulator waveguide is very promising for future quantum networks. With great advances in chip scale quantum state engineering and telecom band fiber based quantum memory, our all fiber based entangled source is directly compatible with these systems for realizing more complex computation and communication tasks.

For conclusion, we prepare both polarization and time-bin entangled photon pairs based on DSF working at room temperature. By taking advantage of the state-of-art DWDM technique, our entanglement source is very promising in entanglement distribution over multi-users. The present work fits in the aim of constructing an all fiber based long distance quantum networks, which provides a scalable, practical and compact platform for quantum technology.

**Acknowledgments**
We thank Dr. Shuang Wang for technical supporting in providing high quality unbalanced Michelson interferometers. This work was supported by the National Fundamental Research Program of China (Grant No. 2011CBA00200), the National Natural Science Foundation of China (Grant Nos. 11174271, 61275115, 61435011, 61525504) and the Fundamental Research Funds for the Central Universities.

**References**
[1]D. Bouwmeester, J.-W. Pan, K. Mattle, M. Eibl, H. Weinfurter, and A. Zeilinger, Experimental quantum teleportation. Nature 390, 575-579 (1997).
[2]I. Marcikic, H. de Riedmatten, W. Tittel, H. Zbinden, and N. Gisin, Nature 421, 509- 513(2003).
[3]X.-L.Wang, X.-D. Cai, Z.-E. Su, M.-C. Chen, D. Wu, L. Li, N.-L. Liu, C.-Y. Lu, and J.-W. Pan, Quantum teleportation of multiple degrees of freedom of a single photon. Nature 518, 516-519(2015).
[4] K. Mattle, H. Weinfurter, P. G. Kwiat, and A. Zeilinger, Dense Coding in Experimental Quantum Communication. Phys. Rev. Lett. 76, 4656-4659 (1996).
[5] M. W. Mitchell, J. S. Lundeen, and A. M. Steinberg, Super-resolving phase measurements with a multiphoton entangled state. Nature 429, 161-164 (2004).
[6] T. Nagata, R. Okamoto, J. L. O'Brien, K. Sasaki, S. Takeuchi, Beating the Standard Quantum Limit with Four-Entangled Photons. Science 316, 726-729 (2007).
[7] G. Brida, M. Genovese, and I. R. Berchera, Experimental realization of sub-shot-noise quantum imaging. Nat. Photon. 4, 227-230(2010)。
[8]P. G. Kwait, K. Mattle, H. Weinfurter, A. Zeilinger, A. V. Sergienko, and Y. Shih, New High-Intensity Source of Polarization-Entangled Photon Pairs. Phys. Rev. Lett. 75, 4337-4341 (1995).
[9] Y. Li, Z.-Y. Zhou, D.-S. Ding, and B.-S. Shi, CW-pumped telecom band polarization entangled photon pair generation in a Sagnac interferometer. Opt. Express 23,28792-28800(2015)
[10] A. Kuzmich, W. P. Bowen, A. D. Boozer, A. B

[11]oca, C. W. Chou, L.-M. Duan, and H. J. Kimble, "Generation of nonclassical photon pairs for scalable quantum communication with atomic ensembles," Nature 423, 731-734 (2003).
[12] D. S. Ding, Z. Y. Zhou, B. S. Shi, X. B. Zou, and G. C. Guo, "Generation of non-classical correlated photon pairs via a ladder-type atomic configuration: theory and experiment," Opt. Express 20, 11433–11444 (2012).
[13] X. Li, P. L. Voss, J. E. Sharping, and P. Kumar, Optical-Fiber Source of Polarization-Entangled Photons in the 1550 nm Telecom Band. Phys. Rev. Lett. 94, 053601(2015).
[14] H. Takesue and K. Inoue, Generation of polarization-entangled photon pairs and violation of Bell's inequality using spontaneous four-wave mixing in a fiber loop. Phys. Rev. A 70, 031802 (2004).
[15] H. Takesue and K. Inoue, Generation of 1.5-µm band time-bin entanglement using spontaneous fiber four-wave mixing and planar light-wave circuit interferometers. Phys. Rev. A 72, 041804 (2004).
[16] J. W. Silverstone, D. Bonneau, K. Ohira, N. Suzuki, H. Yoshida, N. Iizuka, M. Ezaki, C. M. Natarajan, M. G. Tanner, R. H. Hadfield, V. Zwiller, G. D. Marshall, J. G. Rarity, J. L. O'Brien and M. G. Thompson, On-chip quantum interference between silicon photon-pair sources. Nat. Photon. 8, 104-108(2014).
[17] C. Reimer, M. Kues, L. Caspani, B. Wetzel, P. Roztocki, M. Clerici, Y. Jestin, M. Ferrera, M. Peccianti, A. Pasquazi, B. E. Little, S. T. Chu, D. J. Moss, and R. Morandotti, Cross-polarized photon-pair generation and bi-chromatically pumped optical parametric oscillation on a chip. Nat. Comunn. 6, 8236(2015).
[18] C. Reimer, M. Kues, P. Roztocki, B. Wetzel, F. Grazioso, B. E. Little, S. T. Chu, T. Johnston, Y. Bromberg, L. Caspani, D. J. Moss, R. Morandotti, Generation of multiphoton entangled quantum states by means of integrated frequency combs. Science 351, 1176-1180(2016).
[19]J. Fulconis, O. Alibart, J. L. O'Brien, W. J. Wadsworth, and J. G. Rarity, Nonclassical Interference and Entanglement Generation Using a Photonic Crystal Fiber Pair Photon Source. Phys. Rev. Lett. 99, 120501(2007).
[20]J.-W. Pan, Z.-B. Chen, C.-Y. Lu, H. Weinfurter, A. Zeilinger, and M. Zukowski, Multiphoton entanglement and interferometry. Rev. of Mod. Phys. 84, 777-838 (2012).
[21] H. J. Kimble, The quantum internet. *Nature* **453**, 1023-1030(2008).
[22] K. F. Lee, J. Chen, C. Liang, X. Li, P. L. Voss, and P. Kumar, Generation of high-purity telecom-band entangled photon pairs in dispersion-shifted fiber. Opt. Lett. 31, 1905-1907(2006).
[23] S. Dong, Q. Zhou, W. Zhang, Y. He, W. Zhang, L. You, Y. Huang, and J. Peng, Energy-time entanglement generation in optical fibers under CW pumping. Opt. Express 22, 359-368 (2014).
[24]J. Jin, E. Saglamyurek, M. G. Puigibert, V. Verma, F. Marsili, S. W. Nam, D. Oblak, and W. Tittel, Telecom-Wavelength Atomic Quantum Memory in Optical Fiber for Heralded Polarization Qubits. Phys. Rev. Lett. 115, 140501(2015).
[25]E. Saglamyurek, J. Jin, V. B. Verma, M. D. Shaw, F. Marsili, S. W. Nam, D. Oblak,

and W. Tittel, Quantum storage of entangled telecom-wavelength photons in an erbium-doped optical fibre. Nat. Photon. 9, 83-87(2015).

[26] E. Saglamyurek, M. G. Puigibert, Q. Zhou, L. Giner, F. Marsili, V. B. Verma, S. W. Nam, L. Oesterling, D. Nippa, D. Oblak, and W. Tittel. A multiplexed light-matter interface for fibre-based quantum networks. Nat. Comunn. 7, 11202 (2015).

[27] J. F. Clauser, M. A. Horne, A. Shimony, R. A. Holt, Proposed experiment to test local hidden-variable theories. *Phys. Rev. Lett.* **23,** 880–884 (1969).

[28] S. Wang, W. Chen, J.-F. Guo, Z.-Q. Yin, H.-W. Li, Z. Zhou, G.-C. Guo, and Z.-F. Han, 2 GHz clock quantum key distribution over 260 km of standard telecom fiber. Opt. Lett. 37, 1008-1010(2012).

[29]S. Wang, Z.-Q. Yin, W. Chen, D.-Y. He, X.-T. Song, H.-W. Li, L.-J. Zhang, Z. Zhou, G.-C. Guo, and Z.-F. Han, Experimental demonstration of a quantum key distribution without signal disturbance monitoring. Nat. Photon. 9, 832-836(2015).

[30] H. Takesue, and Y. Noguchi, Implementation of quantum state tomography for time-bin entangled photon pairs. Opt. Express 17, 10976-10989 (2009).